\documentclass[sn-mathphys,Numbered]{sn-jnl}

\usepackage{graphicx}%
\usepackage{multirow}%
\usepackage{amsmath,amssymb,amsfonts}%
\usepackage{amsthm}%
\usepackage{mathrsfs}%
\usepackage[title]{appendix}%
\usepackage{xcolor}%
\usepackage{textcomp}%
\usepackage{manyfoot}%
\usepackage{booktabs}%
\usepackage{algorithm}%
\usepackage{algorithmicx}%
\usepackage{algpseudocode}%
\usepackage{listings}%
\usepackage{setspace} 
\usepackage{lineno}

\theoremstyle{thmstyleone}%
\theoremstyle{thmstyletwo}%

\theoremstyle{thmstylethree}%

\begin{document}

\title[Approximate Inference for Longitudinal Mechanistic HIV Contact Networks]{Approximate Inference for Longitudinal Mechanistic HIV Contact Networks}


\author*[1]{\fnm{Octavious} \sur{Smiley}}\email{octavioussmiley@gmail.com}

\author[1]{\fnm{Till} \sur{Hoffmann}}\email{thoffmann@hsph.harvard.edu }

\author[1]{\fnm{Jukka-Pekka} \sur{Onnela}}\email{onnela@hsph.harvard.edu }

\affil*[1]{\orgdiv{Biostatistics}, \orgname{Harvard University}, \orgaddress{\street{677 Huntington Ave}, \city{Boston}, \postcode{02115}, \state{MA}, \country{USA}}}




\abstract{\doublespacing Network models are increasingly used to study infectious disease spread. Exponential Random Graph models have a history in this area, with scalable inference methods now available. An alternative approach uses mechanistic network models. Mechanistic network models directly capture individual behaviors, making them suitable for studying sexually transmitted diseases. Combining mechanistic models with Approximate Bayesian Computation allows flexible modeling using domain-specific interaction rules among agents, avoiding network model oversimplifications. These models are ideal for longitudinal settings as they explicitly incorporate network evolution over time. We implemented a discrete-time version of a previously published continuous-time model of evolving contact networks for men who have sex with men (MSM) and proposed an ABC-based approximate inference scheme for it. As expected, we found that a two-wave longitudinal study design improves the accuracy of inference compared to a cross-sectional design. However, the gains in precision in collecting data twice, up to 18\%, depend on the spacing of the two waves and are sensitive to the choice of summary statistics. In addition to methodological developments, our results inform the design of future longitudinal network studies in sexually transmitted diseases, specifically in terms of what data to collect from participants and when to do so.}





\maketitle

\section{Introduction}

Networks are used to study a range of systems with interactions or dependencies among their agents, such as the behavior in supply chains and the stock market \citep{Macal}, protein-protein interactions in biological systems \citep{Scholtens}, and disease transmission on local and global scales \citep{Le}. In the study of disease transmission dynamics, the contact structure of a population can be naturally represented as a network, and this representation is especially useful if the contacts persist over time, as is often the case for sexual interactions. Disease dynamics are then driven by interactions (represented by edges) among susceptible and infectious individuals (represented by nodes). More generally, many of these systems arise from stochastic processes forming or dissolving interactions over time that must be accounted for when doing inference.

There are (at least) two main paradigms of networks models: statistical and mechanistic. Statistical network models prioritize tractable likelihoods to facilitate inference at the expense of model flexibility. For example, the Erd\H{o}s\textendash{}R\'{e}nyi  graph, also known as the Bernoulli random graph, assumes that each node pair is connected independently and with identical probability. Hence, the likelihood of the number of edges is the standard binomial likelihood with a fixed number of nodes and inference readily follows because a graph is completely identified by its node and edge sets. It also follows that Erd\H{o}s\textendash{}R\'{e}nyi has a binomial (approximately Poisson) degree distribution.

This generative mechanism however clearly does not map well to most real-world networks. This is easily seen in the World Wide Web (WWW). In this scenario, each website is represented by a node and a directed connection (hyperlink) between websites occurs when one links to the other. Unlike Erd\H{o}s\textendash{}R\'{e}nyi  networks, where the degree distribution follows a binomial distribution, the degree distribution here follows a power-law where more successful websites tend to grow their connections faster than others \citep{WWW}. Exponential Random Graph Models (ERGMs) are generalizations of the Erd\H{o}s\textendash{}R\'{e}nyi model. They represent a probability distribution of graphs on a fixed node set, where the probability of observing a graph is dependent on the presence of the various configurations specified by the model \citep{ERGMs}. A typical graph in this distribution can be interpreted as the aggregate of the local configurations, and slight errors in estimating the local configuration counts can alter beliefs about the distribution \citep{ERGMs_Onnela}.

Mechanistic models assume that the observed network is generated by a small set of mechanistic rules. 
The canonical example is the Barab\'{a}si-Albert (BA) model. Nodes are added one by one to a growing network and each node connects to $m$ previously existing nodes with probability proportional to the nodes' current degree \citep{BA}. This so-called preferential attachment mechanism readily generates power-law degree distributions, which are a type of broad-tailed degree distribution that are characteristic of many empirical networks, including that of the WWW. Apart from the target number of nodes, $n$, the classic BA model only has one free parameter, $m$. In this case, the fully grown graph has approximately $nm$ edges, and $m$ can be inferred by dividing the number of edges by the number of nodes $n$, i.e., $m$ is approximately equal to the average degree of the network. However, even for moderately complex models, the likelihood of the full network becomes intractable due to the fact that the insertion order of the nodes is (usually) not known. Because the graph is sequentially dependent on the previous iteration as it grows, the number of possible graph realizations grows exponentially with the number of added nodes.

Networks have provided insights to major public health problems such as the spread of HIV \citep{victor}, the opioid crisis \citep{Opiod_ERGMS}, and interventions with people who inject drugs \citep{PWID}. Wertheim et al.~noted that HIV is an evolving disease and constructed a disease transmission network using gene sequencing by tracking the evolutionary path of the virus and inferring edges by measuring the similarity of the virus within different individuals. Using an inferred transmission network, they developed a network statistic that was able to detect community level effects of HIV in a clinical trial setting that could help thwart future infections \citep{victor}. Aroke et al.~showed the benefit of peer influence and concluded that individuals who have a diagnosis of opioid use disorder or use many prescribers may help promote positive health behaviors in an opioid prescription network due to the influence of their direct peers on the network structure. They came to this conclusion by showing the type of opioid that an individual uses and their number of prescribers were identified as significant predictors of high betweenness centrality giving them influence over the network at large \citep{Opiod_ERGMS}. Rolls et al.~model network data involving people who inject drugs, using validation techniques, so that these networks can be simulated and intervention strategies could be explored \citep{PWID}.

There are several mechanistic models for studying the impact of men who have sex with men (MSM) contact networks and their impact on HIV transmission \citep{chicago, Mei, Hansson}. Birkett et al.~used a data-driven simulation model to understand the impact of network-level mechanisms and STI infections on the spread of HIV among Young Men who have Sex with Men (YMSM) \citep{chicago}. Mei et al.~introduced the concept of a Complex Agent Network (CAN) to model the HIV epidemics by combining agent-based modelling and complex networks \citep{Mei}. An especially interesting model was introduced by Hansson et al.~to study the role of casual contacts on the HIV epidemic in Stockholm, Sweden. Their research was used to recommend interventions to reduce transmission rates \citep{Hansson}. Padeniya et al.~notes Hansson and others' contribution to intervention strategies as they sought to mathematically model the role of female-sex-worker-client interactions for gonorrhoea transmission \citep{Padeniya}. Vajdi et al.~noted Hansson's choice to model instantaneous casual relationships, and investigated a dynamic model for casual relationships, a two-layer temporal network model, and SIS mean-field equations \citep{Vajdi}. A common approach for inference in these papers is to propose mechanisms for contact formation, simulate the spread of disease on the network, and modify parameter values to match disease prevalence to that observed in their respective populations without directly validating their mechanism. 

Most scientific studies involving human subjects can be divided into cross-sectional and longitudinal. In cross-sectional studies, measurements are obtained at only a single point in time. The distinguishing feature of longitudinal studies is that the study participants are measured repeatedly (at least twice) throughout the duration of the study, thereby permitting the direct assessment of changes in the response variable over time \citep{ALA}. To illustrate, participants in a cross-sectional study likely vary in age; however, this type of design cannot be used to study the effect of aging because the effect of aging is potentially confounded with cohort effects. 
It is important to note here that although we are sampling the evolving network at multiple time points, we are only asking participant information that can maintain privacy. 

One example of a longitudinal network study is the work by Birkett et al.~\citep{chicago}. The authors studied the impact of network-level mechanisms and STI infections on the spread of HIV and found that network-level mechanisms and STI infections play a significant role in the spread of HIV and in racial disparities among (YMSM). Their work shows HIV prevention efforts should target YMSM across race, and interventions focusing on YMSM partnerships with older MSM might be highly effective. In general, one would expect observing a network multiple times to provide more information, and therefore improve accuracy of inference, compared to observing the network just once. In addition, one may address questions that can only be interrogated in a longitudinal study. When growing a network in a simulation, we can track every iteration of the dynamic network and have arbitrarily many observations at our disposal. In an actual study, one is of course constrained by resources and logistics. If the data are obtained from self-administered or staff-administered surveys, too frequent reporting may lead to participant burden and reduce his or her willingness to continue participation, whereas too infrequent reporting may lead to recall bias and participants may be lost to follow up. For example, a person may not remember each individual whom they dated over a five-year period and may not be able to reliably recall the timing of the relationships. Collecting data at different time points that are optimally spaced helps alleviate recall bias  while still maintaining an avenue for accurate inference.

Our goal in this paper is to implement a discrete-time version of the mechanistic network model introduced by Hansson et al.~and use the model to identify optimal spacing between two data collection points (waves) in a longitudinal network study such that we can achieve the dual goal of accurate inference (learning model parameters as precisely as possible) while minimizing participant burden (using network features that in practice could be elicited from participants with a minimal number of survey questions). These results have implications for study design for HIV and other sexually transmitted diseases, and more broadly they can inform other research questions involving (longitudinal) network data.  

This paper is structured as follows. We discuss the discrete-time mechanistic network model in Section \ref{sec:mechmod} and explain our ABC-based approach to approximate parameter inference in Section \ref{sec:inference}. We show our results in Section \ref{sec:results} and conclude with a discussion in Section \ref{sec:discussion}.

\section{Methods}

\subsection{Mechanistic network model}
\label{sec:mechmod}
As noted in the Introduction, there are several mechanistic models for MSM contact formation in specific populations. We focus on the mechanistic model introduced to study MSM contact networks in Stockholm, Sweden \citep{Hansson}. The model incorporates specific behaviors that guide the formation and dissolution of sexual contacts as well as migration of individuals in and out of the population. While the original model was formulated in continuous time, we consider a discrete time version of the model. This means that rates in the original formulation correspond to probabilities in ours. We note that as the number of the potential discrete time events tends to infinity and the event probabilities tend to zero, our formulation of the model converges to the original. Throughout this paper, each discrete model time step iteration is taken to correspond to one calendar month, and all events are recorded at the end of each iteration. While a constant number of individuals enter the population at each iteration, each individual leaves the population with a fixed probability at each iteration. The size of the network therefore fluctuates around $n$ nodes, where $n$ is the initial number of nodes in the network.

The model incorporates two types of partnerships: steady and casual. Casual relationships are defined to only last one iteration at onset while steady relationships are defined to have the potential to last longer. An individual can have at most one steady partner at any given time. The probability that a single person enters a steady relationship at a given iteration is $\rho P_0$, where $P_0$ is the proportion of single individuals in the present iteration. In the original model, where $\rho$ is a rate of steady partnership formation, $P_0$ fluctuates around an equilibrium; in our version, we fix this parameter and absorb it into $\rho$ for simplicity and to improve identification of model parameters. Our modified probability of a single person entering a steady relationship at a given iteration is therefore $\rho$. While the number of people willing to form relationships varies from iteration to iteration, the probability a single person joining a relationship stays the same. In the Hansson paper, the differential equation formulation of the model explicitly considers the fluctuation of the likelihood of new relationships while we do it implicitly as the number of singles changes. The probability of said steady relationship dissolving at each iteration is $\sigma$.

\begin{figure}
\includegraphics[width=\textwidth]
        {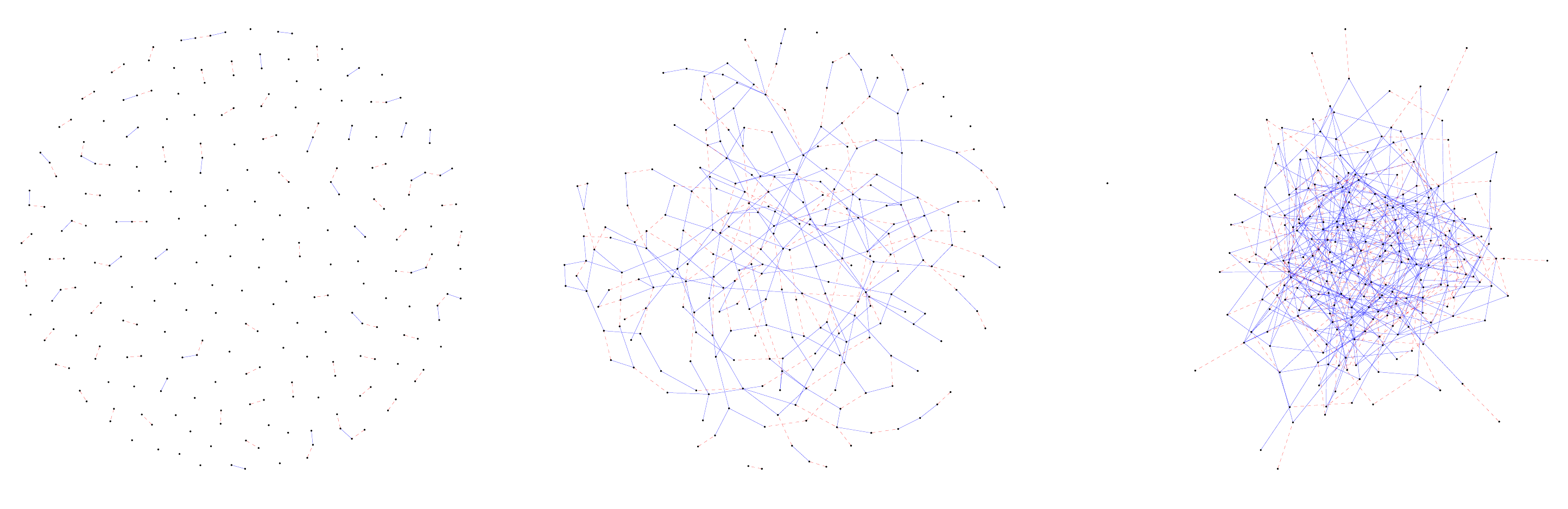}
\caption{Network visualizations containing cumulative (from iteration 1) steady (red dashed) and casual (blue solid) edges for iterations 1 (left), 6 (middle), and 12 (right). We used the following parameter values: $\mu$ = 0, $\rho$ = 0.3, $\sigma$ = 0.1, $w_{1}$ = 0.2, $w_{0}$ = 0.4.}
\label{network_vis}
\end{figure}

In addition to a steady relationship, an individual may also have one casual partnership at each iteration. These casual relationships may occur alongside steady partnerships or during times when the person is single. A single individual enters a casual relationship with probability $\omega_0$ while an individual who is currently in a steady relationship forms a casual relationship with probability $\omega_1$. For any partnership to form, both individuals must be willing to join that relationship. In the scenario where an odd number of individuals would like to form a relationship, one of them (chosen at random) is left out. Each person migrates from the population with probability $\mu$, and individuals enter into the population at constant rate $n\mu$. The migration of an individual and the formation and dissolution of a sexual contacts are all determined by the outcome of independent Bernoulli trials. In the original continuous time formulation of the model, duration of steady relationships and the time spent in the population both follow exponential distributions. In contrast, for our discrete time formulation both are geometrically distributed. Starting from an empty graph with $n$ nodes, we first run the model until we are confident that it has converged to the target distribution. We set our migration probability to 0 to ensure we are sampling individuals longitudinally and to maintain a closed cohort design. We note that the 'constant' number of nodes being added is largely dependent on only $\mu$ and $n$ and easily recoverable. The model is described in Algorithm \ref{alg} and a few graph realizations from the model are illustrated in Figure \ref{network_vis}. We chose 1000 iterations to ensure we are past the burn-in \citep{Hansson}.

\begin{algorithm}
\caption{Hansson MSM model \citep{Hansson}}
\label{alg}
\begin{algorithmic}[1]
\State \textbf{Inputs}:
\State $n$ := number of nodes
\State G := (V, E) {the graph has no edges}
\State $\rho$ := scaling parameter for partnership formation probability
\State $\sigma$ := separation probability for steady relationships
\State $\omega_0$ := the probability someone who is single enters into a casual relationship
\State $\omega_1$ := the probability someone who is in a partnership enters into a casual relationship
\State $iterations$ := the number of iterations to run algorithm

\State \textbf{Algorithm}:

\For{i in 1:iterations}
    \State Dissolve all casual relationships generated from the previous iteration if applicable
    \State Identify all current steady relationships
    \If{there are a positive number of steady relationships}
        \State Dissolve each with probability $\sigma$
    \EndIf
    \State Identify all nodes with degree 0
    \State Set willingness to form a steady relationship with probability $\rho$
    \State Randomly match the maximum number of even nodes that are willing to form a steady relationship into edges
    \State Identify single nodes := nodes with degree 0
    \State Set willingness to form a casual relationship with probability $\omega_0$
    \State Identify steady nodes := nodes with degree 1
    \State Set willingness to form a casual relationship with probability $\omega_1$
    \State Randomly match the maximum number of even nodes among the single and steady nodes together that are willing to form a casual relationship
    \State Record Network
\EndFor
\end{algorithmic}
\end{algorithm}

\subsection{Inference of model parameters}
\label{sec:inference}
\begin{figure}
\includegraphics[width=\textwidth, height = 0.66\textwidth]{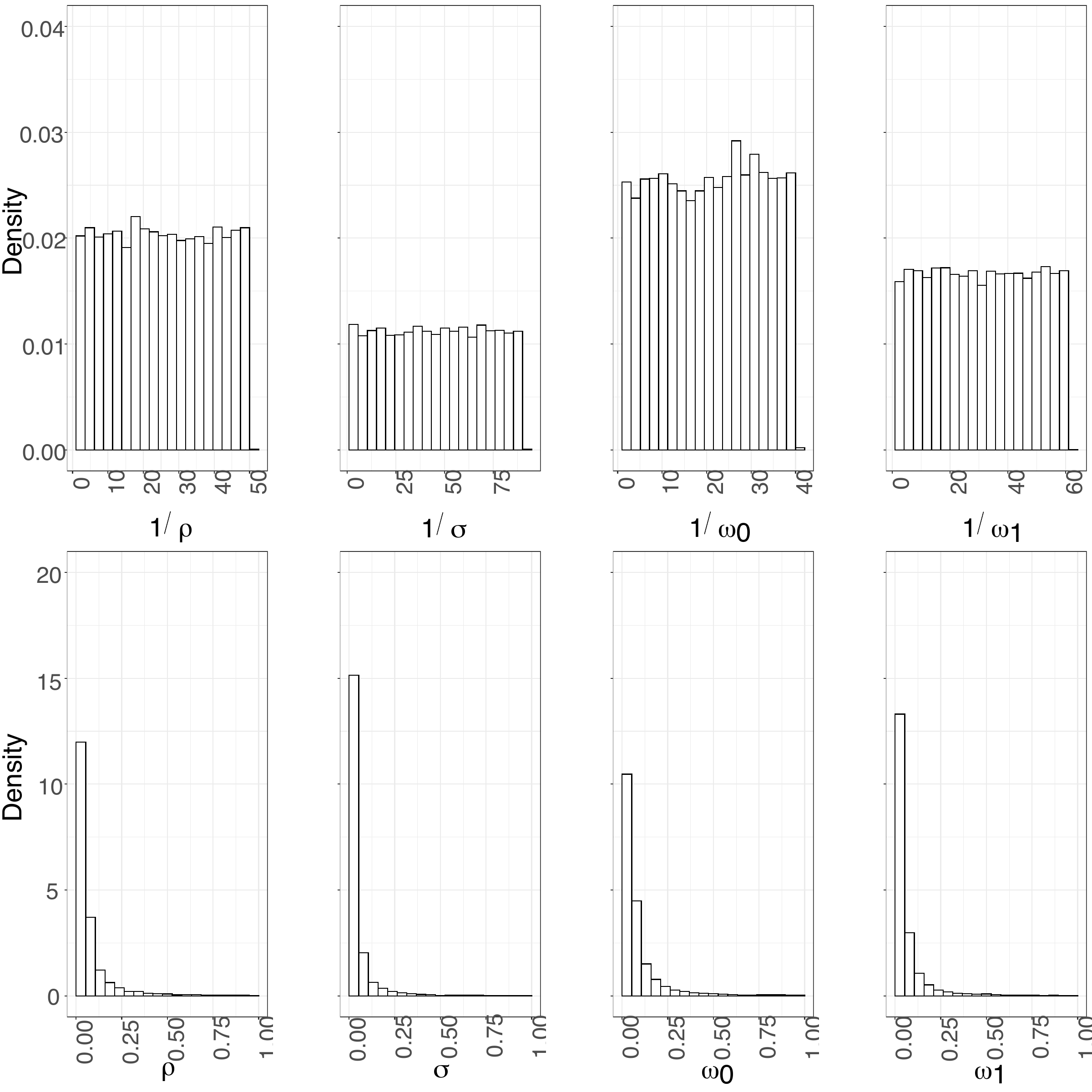}
\caption{Prior distributions on the inverse parameters and the corresponding implied prior distributions on parameters themselves. The top row shows the distributions of the inverse parameters, which can be interpreted as distributions of the average values of geometric distributions. The bottom row shows the distributions of the parameter values themselves for our discrete time mechanistic network model for the following parameters: $\rho$, $\sigma$, $\omega_1$, $\omega_0$.}
\label{prior_plots}
\end{figure}

In Bayesian inference, complete knowledge of the model parameters, given the observed data, is contained in the posterior distribution. Typically, in mechanistic models, the complexity of the model means that the likelihood and corresponding posterior distribution is not available in closed form. In mechanistic models one can nevertheless forward simulate data from the model given parameters, and these parameter values may be obtained from a prior distribution. ABC is an inference framework that has been developed to deal with models that have intractable likelihoods. There are several ABC methods to generate samples from an approximate posterior distribution. For clarity of our objective, we use the simple accept/reject algorithm. In accept/reject, we propose parameter values from a prior distribution to generate data and retain parameter values that produce data that resembles the observed data \citep{csillery}. If we only kept parameter values that reproduced the observed data exactly, this approach would recover the exact posterior for discrete data. This approach is however impossible for continuous data because the probability of sampling a continuous value exactly is 0. To ensure that our prior distribution's support is realistic to MSM relationship characteristics, we utilize a uniform distribution on the duration of average time spent for people to be open to joining a relationship $\frac{1}{\rho}$ [1 month, 50 months], average time a steady relationship lasts $\frac{1}{\sigma}$ [1 month, 90 months], average time for a single individual to partake in a casual relationship $\frac{1}{\omega_0}$ [1 month, 40 months], and average time for an individual in a relationship to partake in a casual relationship $\frac{1}{\omega_1}$ [1 month, 61 months]. Figure \ref{prior_plots} shows the prior distributions on these inverse parameters and the corresponding implied prior distributions on the parameters themselves. We recognize a variety of definitions for steady and casual relationships in MSM contact networks, as well as a variety of estimates for the support of each duration \citep{malone, down2017, de2000, wall, davidovich, weiss, bavinton, myers}. We chose our support to be consistent with the data the model was originally trained on \citep{Hansson}, and calculate the reciprocal of each parameter sampled from the prior as an input to our model. 

There are at least three major considerations in the ABC accept/reject framework: summary statistics, distance measure, and similarity threshold \citep{Sisson}. Given the mechanistic network model of interest, we manually chose a set of summary statistics needed for inference, which renders the network space more manageable \citep{Sisson}. The choice of network summary statistics was guided by the principle that it should be possible to obtain this information from study participants using a questionnaire and they should be informative of the model parameters. At a minimum, one needs at least as many summary statistics as there are parameters to be inferred \citep{Sisson}. Although the model has five parameters (six if one counts $n$), as previously mentioned, we opted to fix one of them, the migration probability $\mu=0$. This leaves us with four parameters to recover: probability of a single person entering a steady relationship $\rho$, probability of dissolving a steady relationship $\sigma$, probability of a single individual to enter a casual relationship $\omega_0$, and probability of an individual in steady relationship to enter a casual relationship $\omega_1$. More information on parameters can be found in Table \ref{param_interp}. We chose the four summaries, denoted $s_1$ through $s_4$, as listed in Table \ref{summ_description}. We chose summaries that could be elicited by asking participants to consider their sexual history in the past year only. Longer histories could potentially be more informative, but longer look-back periods would likely increase recall bias.

Since the mechanistic network model is outside the exponential family, we have no guarantee of sufficiency, i.e., that our summary statistics fully summarize our network. However, we still require the summary statistics to be informative of the model parameters. An informal way to assess the extent of informativeness is to investigate plots of network summary statistics against model parameters. We denote these relationships as $s_i(\theta)$ where $i \in \{1, 2, 3, 4\}$ and $\theta \in \{\rho, \sigma, w_0, w_1 \}$. We refer to these relationships as \emph{mapping functions}, and we estimate them using simulations where $s_{i, k}(\theta)$ represents the value of summary statistic $s_i$ with respect to generative parameter $\theta$ in simulation run $k$. The value of $s_i(\theta)$ is given as the median value of $s_{i, k}(\theta)$ taken across all simulations $k$.

We measured the distance between a simulated network and the observed network by calculating the Euclidean distance within the normalized summary statistic space. The normalized summary statistic value is obtained by first subtracting the mean of the summary statistic from each value and then dividing each value by the standard deviation of the summary statistic. We populated an ABC reference table for each lag by generating 10,000 graphs by sampling parameters from their joint priors and varying the lag between the two network observations between zero and 150 iterations. Next, taking a sample per lag from our joint prior density and its corresponding graph as our ground truth, we simulated samples from the corresponding approximate posterior distribution. We retained the parameters associated with the 100 (top 1\%) smallest distances in the normalized summary statistic space between the observed and generated graphs.

Finally, we performed a regression adjustment on samples from the approximate posterior distribution \citep{Beaumont, Beaumont2}. The goal of the regression adjustment is to improve our ABC posterior's convergence to our target posterior. The basis of the method is that we can obtain an estimate of our expected parameter values given the summaries using linear regression in the localized neighborhood around our observed data that we get from the approximate posterior. Then, we can use this relationship to adjust our approximate posterior distribution \citep{Beaumont}. We normalized the parameters in our reference table, and utilized the root mean squared error (RMSE) of the approximated posteriors for a fixed set of 500 ground truth parameters to measure accuracy of inference. Then, we averaged over all 500 parameter sets for an estimate of the RMSE, for a given lag, over our prior space \citep{Fearnhead}. For clarity, consider $\theta_{i}$ as the $ith$ ground truth parameter and $\hat{\theta}_{i,k}$ as the $kth$ sample from the approximate posterior estimating $\theta_{i}$. Our estimate of RMSE is then
\begin{equation}
    \hat{RMSE} = \frac{1}{500} \Sigma_{i = 1}^{500} \sqrt{\frac{1}{100} \Sigma_{k = 1}^{100} (\theta_{i} - \hat{\theta}_{i, k})^2}.
\end{equation}

Finally, we fitted a locally weighted regression (loess) with a 95\% confidence interval to the data. 
We note that while the values of the parameters in the reference table are normalized, the resulting approximate posterior distributions are not. Individually normalizing the reference table parameters is useful because it places all parameters on the same scale when calculating the RMSE. However, the posteriors are displayed on the original scale for ease of interpretation.

\begin{table}
    \centering
    \begin{tabular}{p{.2\columnwidth}p{.4\columnwidth}p{.5\columnwidth}}
    \toprule
         Parameter & Hansson et. al \citep{Hansson} & Current Paper \\
    \midrule                                                   \\
n         & Average population size   & Population size    \\           
$\mu$        & Rate of leaving the sexually active population                     & Probability of leaving the sexually active population            \\
$\rho$       & Partnership desire scale rate                                         & Probability of a single person entering a steady relationship                        \\
$\sigma$     & Separation rate                                                    & Probability of dissolving a steady relationship                                          \\
$\omega_0$      & Rate at which an individual who is single tries to have casual sex & Probability of a single individual to enter a casual relationship \\
$\omega_1$      & Rate at which an individual who is in a relationship tries to have casual sex  & Probability of an individual in a steady relationship to enter a casual relationship\\
\bottomrule
\end{tabular}
\caption{Parameter And Their Interpretation}
    \label{param_interp}
\end{table}

\begin{table}[]
\centering
\begin{tabular}{p{.1\columnwidth}p{.9\columnwidth}}
    \toprule
         Name & Description \\
    \midrule
$s_1$     & The proportion of single individuals   \\
$s_2$ & The average length of steady relationships that start and end in the course of the study \\
$s_3$ & The proportion of individuals in steady relationships who are also in  casual relationships  \\
$s_4$       & The proportion of steady relationships among all relationships\\
\bottomrule
\end{tabular}
\caption{Summary Statistic Descriptions}
    \label{summ_description}
\end{table}

\section{Results}
\label{sec:results}
\begin{figure}
\includegraphics[width=\textwidth]
        {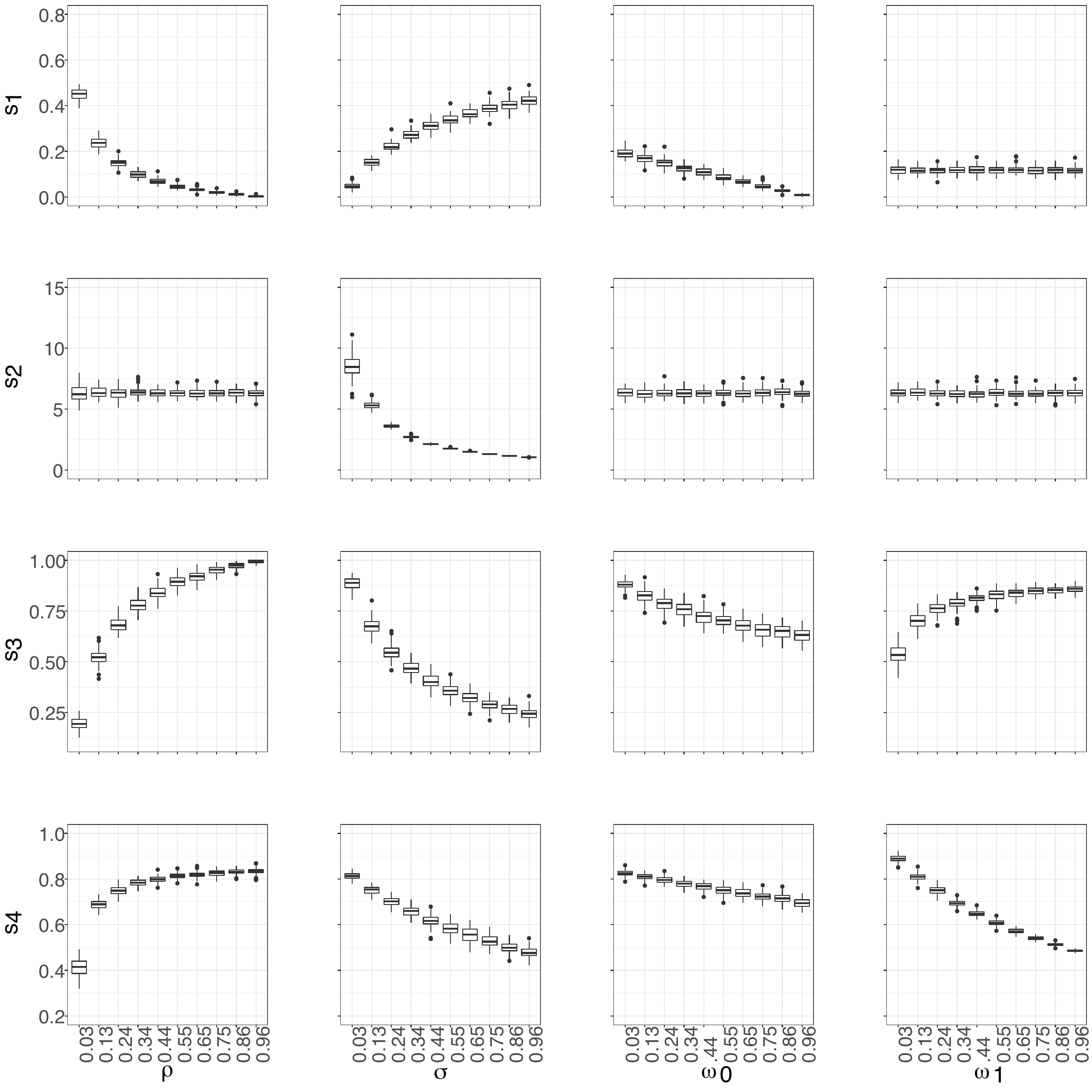}
\caption{Pairwise relationships between the model parameters (horizontal axes) and the summary statistics (vertical axes) used in our ABC inference scheme. Free parameters are fixed at $\mu$ = 0, $\rho$ = 0.3, $\sigma$ = 0.1, $\omega_0$ = 0.4, $\omega_1$ = 0.2. The lag between two consecutive network observations is fixed at 15 iterations. Each box plot consists of 100 samples.}
\label{mapp_fixed}
\end{figure}

\begin{figure}
\includegraphics[width=\textwidth]
        {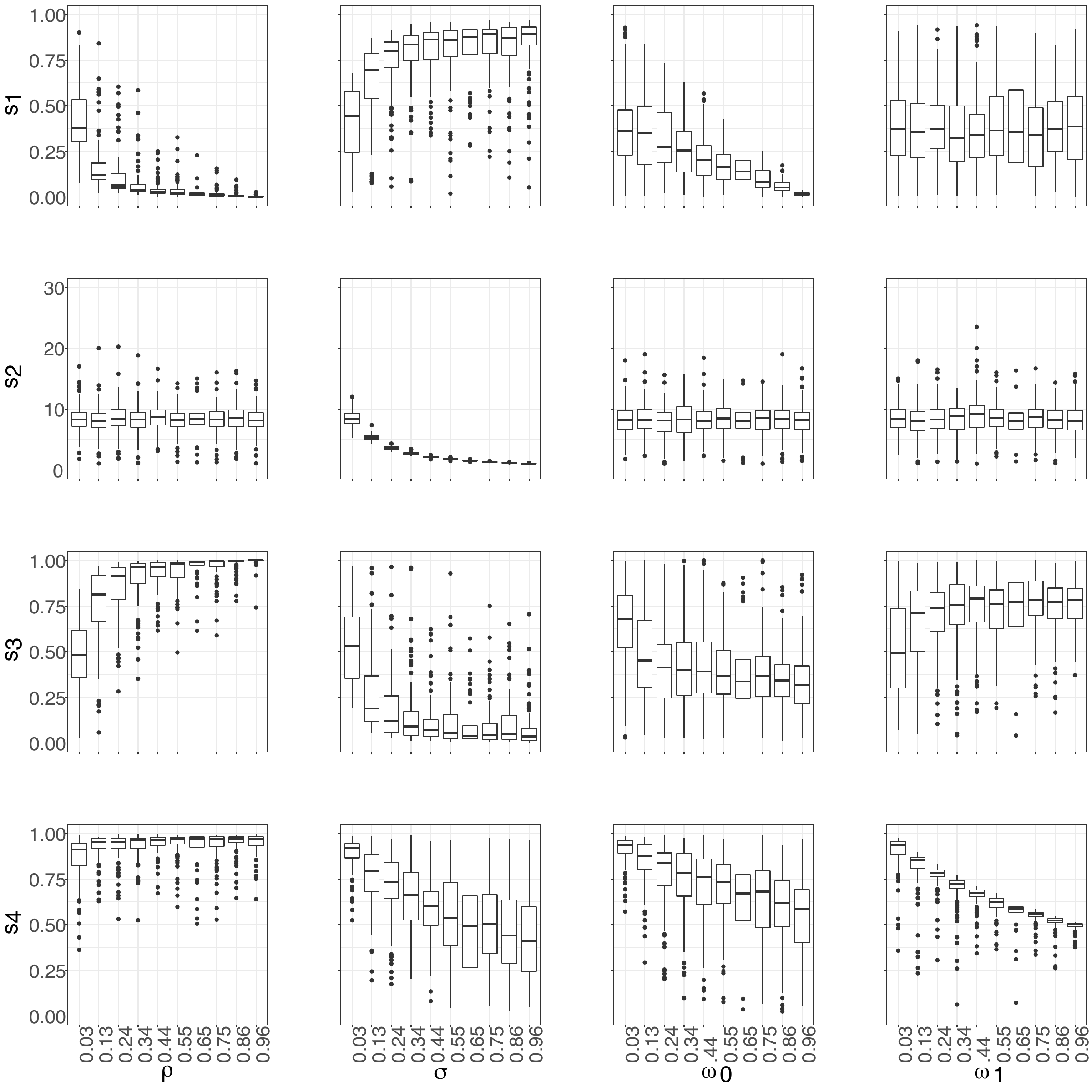}
\caption{Pairwise relationships between the model parameters (horizontal axes) and the summary statistics (vertical axes) used in our ABC inference scheme. Free parameters are sampled from the prior distributions. The lag between two consecutive network observations is fixed at 15 iterations. Each box plot consists of 100 samples.}
\label{mapp_prior}
\end{figure}

We evaluated the mapping functions on a grid along the unit interval by generating 100 graphs per parameter value and using box plots to summarize the results.
We investigated mapping functions in two different scenarios. First, we varied each parameter in turn while keeping all others fixed at the values reported in \citep{Hansson}, i.e., we fixed $\rho$ = 0.3, $\sigma$ = 0.1, $\omega_0$ = 0.4, and $\omega_1$ = 0.2, and we also set $\mu$ = 0 to keep the cohort closed. Second, we sampled each free parameter from its respective prior distribution. These plots were used to ensure that the chosen summary statistics are informative about the model parameters as can be seen in Figures \ref{mapp_fixed} and \ref{mapp_prior}. 

While more summaries could be included, that would increase the computational burden and likely would not significantly increase accuracy. We considered several extra summaries during discovery. Since we would like to obtain the data from questionnaires, one also needs to consider participant burden: all else equal, we would like to ask as few questions as needed to address the scientific question at hand. It is also worth emphasizing that each of the listed summaries can be obtained using privacy preserving questions only in data collection, i.e., participants do not need to disclose their identity nor the identity of their steady or casual partners. This arguably improves the quality of the collected data as respondents would be expected to be more likely to report their behavior accurately. 
We note that while a regression adjustment generally improves the results, it can at times generate functionally impossible values, such as negative probabilities, or worsen our inference when the summaries do not accurately represent the network. In the rare occasion the regression adjustment proposes a negative number, we opt to take a conservative approach and set the value at 0. In this study, we did not see any adjusted proposal probabilities above 1.

\begin{figure}
\includegraphics[width=\textwidth]
        {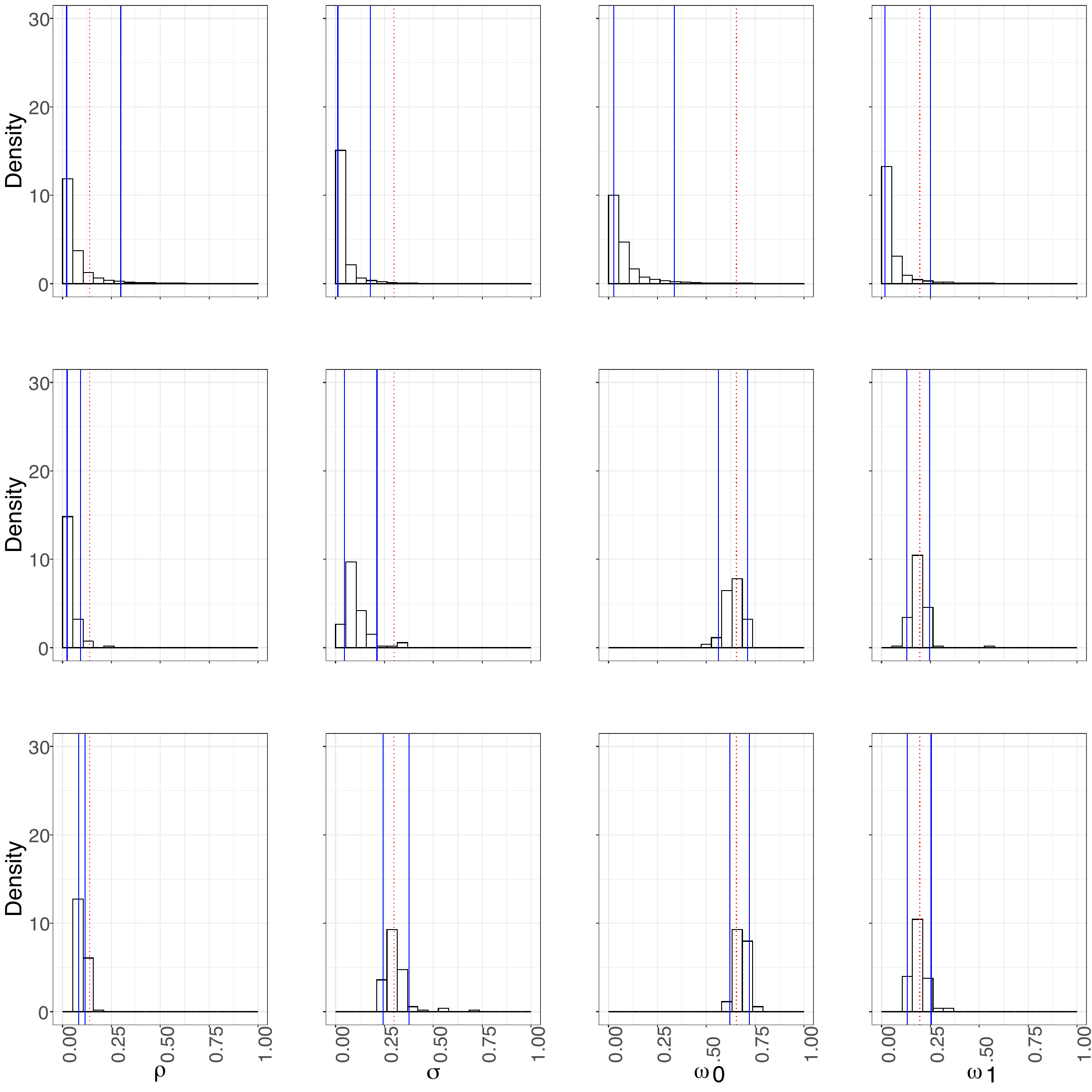}
\caption{Approximate marginal posterior distributions of model parameters obtained by retaining the top 1\% of proposed prior samples in our ABC accept/reject inference scheme. Different rows correspond to comparing the prior (top), observing the graph once (middle), and observing the graph twice with a lag of 50 iterations (bottom). All posteriors include a regression adjustment. The blue solid lines represent the 95\% credible intervals and the red dotted lines represent the true parameter values.}
\label{post_plots}
\end{figure}

\begin{figure}
\includegraphics[width=\textwidth]
        {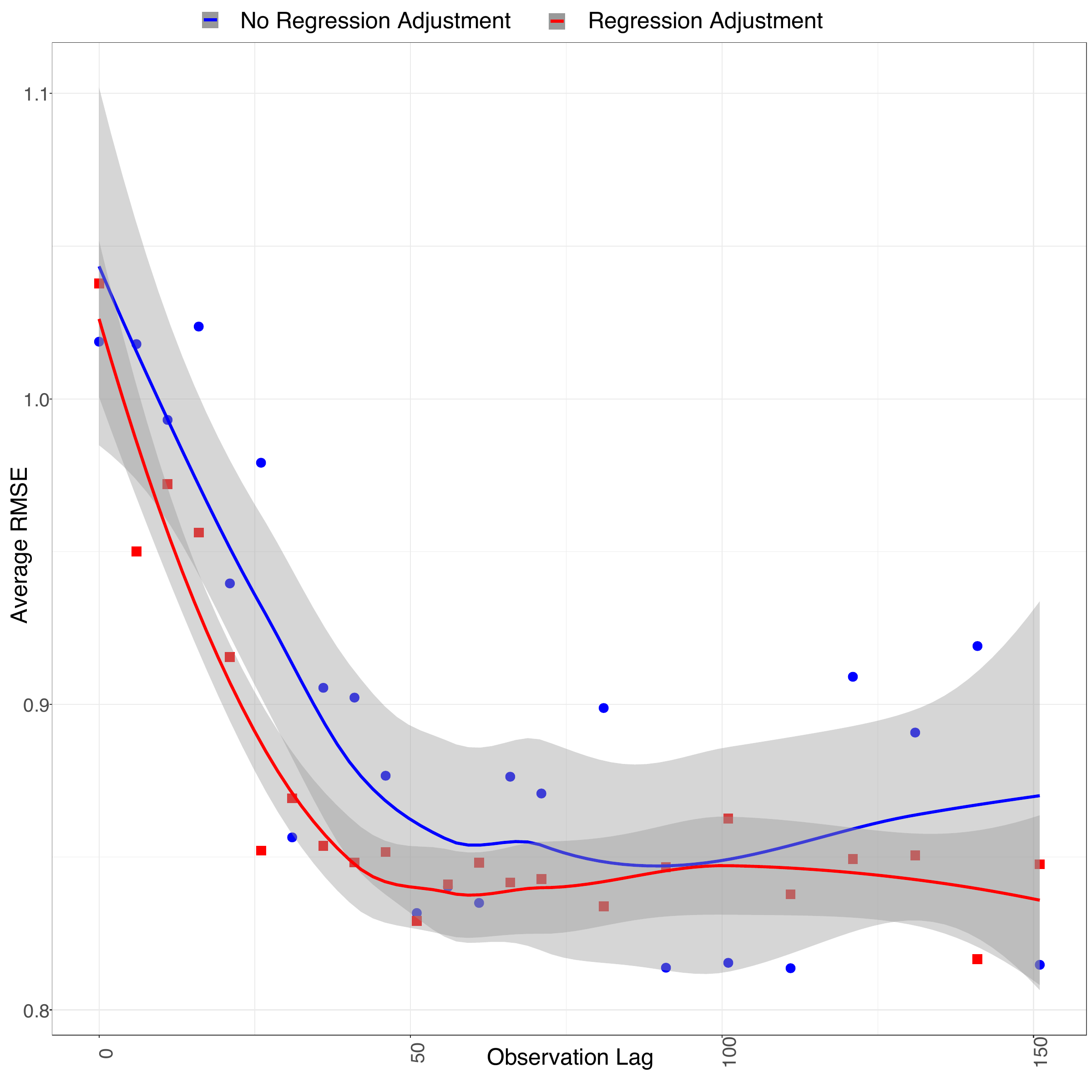}
\caption{Estimated average RMSE, where the average is taken across multiple network realizations, as a function of the lag between the two network observations. We also include a loess curve with a 95\% confidence interval (shaded areas). The average prior average RMSE is $2.22$ (not shown), whereas the corresponding regression adjusted error for a network observed only once is $1.03$ that for a network observed twice with a lag of 50 iterations is $0.84$.}

\label{RMSE_Plots_NoEL}
\end{figure}

\begin{figure}
\includegraphics[width=\textwidth]
        {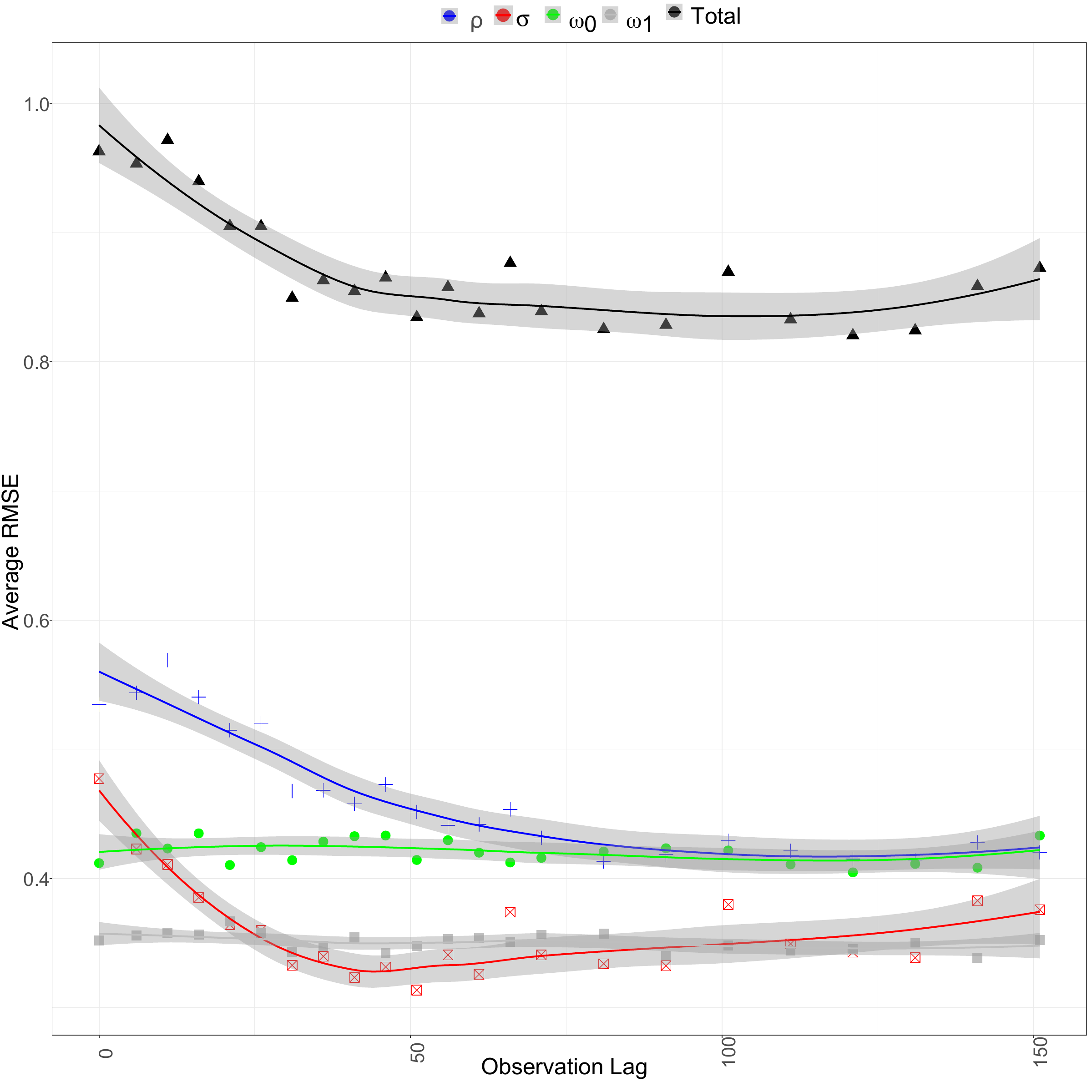}
\caption{Estimated regression adjusted average RMSE for the total error (top curve) and separately for the four parameters considered in our study (bottom four curves). These results show that when observing a network twice, the reduction in total RMSE is mainly due to the reduction of RMSE for $\rho$ and $\sigma$.}
\label{RMSE_parameter_plots}
\end{figure}

We visualized the regression adjusted approximate posteriors when looking at the graph once or twice with a lag of 50 iterations in Figure \ref{post_plots}. Furthermore, as expected, and as shown in Figure \ref{RMSE_Plots_NoEL}, observing a network twice results in a smaller average error compared to observing a network only once. The improvement is largely driven by our ability to recover $\sigma$ and $\rho$ parameters as shown in Figure \ref{RMSE_parameter_plots}. We also see the average error steadily decreases with the lag between the two network observations until about 40 to 50 iterations. This lag between the two network observations (data collection waves) is optimal in the sense that extending the gap further does not greatly increase accuracy of inference but does lengthen the duration of the study. In a closed cohort study, all else equal, the longer the duration of the study, the greater the expected attrition of study participants. Attrition of study participants in a setting like ours would lead to incomplete ascertainment of network structure and therefore introduce an additional error to network summary statistics. We also see that implementing a regression adjustment does reduce our average error by nearly an additional 2.6\%, while maintaining the overall trend and optimal lag. Finally, we note our overall ability to discern parameters from our joint prior distribution when collecting data twice after a regression adjustment with an optimal drop of roughly 62\% from our average prior error and 18\% when only collecting data once.

\section{Discussion}
\label{sec:discussion}

In this paper, we investigated the accuracy of an approximate inference scheme applied to an evolving mechanistic network model in a setting where the network, representing sexual contacts among people in a closed population, is observed at two different time points. As expected, observing the network twice improves the accuracy of inference, but this reduction in inferential error depends on the time lag between the two observations. Given that collection of real-world sexual network data is expensive and logistically challenging, it pays off to optimize the gap between the two time points to maximize accuracy of inference. If the two network observations are too close in time, there may have been only minimal changes in the network structure, and therefore the second observation adds little information. However, if the two network observations are too far apart in time, the study may be logistically difficult to carry out in practice and the population is likely to experience significant churn.

There are a total of six parameters in the model, but we fixed two of them to focus on a closed, fixed-sized cohort. When considering the contribution of the remaining four parameters to inferential error, we observed that the $\sigma$ (probability of dissolving a steady relationship) and $\rho$ (probability of a single person entering a steady relationship) parameters benefited the most from the lag between the two network observations. This finding is intuitive as these two parameters influence multiple relationship iterations. However, $\omega_{0}$ (probability of a single individual to enter a casual relationship) and $\omega_{1}$ (probability of an individual in steady relationship to enter a casual relationship) both correspond to one-time events and do not benefit as much from a lag. In particular, $\omega_{1}$ is relatively accurate at all lags while $\omega_{0}$ would likely see more relative improvement through the consideration of another summary statistic.

        
The set of summary statistics that may be considered in inference depends on the information obtained from subjects through study questionnaires. The informativeness of questions themselves depends on the mechanisms that drive contact formation in the study population. Depending on the mechanisms, it is possible that any set of individual-level questions (giving rise to so-called egocentric samples of the network) may be inadequate for network inference and instead one may need information about the full network structure. While this type of network-level information could be obtained using a sociocentric design, it is very challenging, and we are aware of only one study that has implemented this in practice. The Likoma Network Study was based on a sociocentric survey of sexual partnerships aimed to investigate the population-level structure of sexual networks connecting the young adult population of several villages on Likoma Island, Malawi \citep{helleringer}. We stress that this notable study is cross-sectional and therefore corresponds to a one-time observation of the network (even if the data collection in this study occurred in two stages for logistical reasons). Obtaining two observations of the network would be logistically nearly impossible, and doing so in larger populations is not feasible.

Our results highlight the importance of using simulation to investigate the hypothesized generative mechanisms of network formation to inform future study designs, here specifically 1) what questions to ask so that maximally informative network summary statistics may be constructed and 2) how to space the two (or possibly more) data collection waves. For example, in our setting, introducing extensive migration in the population leads to a shorter optimal lag between the two network observations. Our approach is compatible with the recommended paradigm of using simulations for designing and interpreting intervention trials in infectious diseases, particularly with regard to emerging infectious diseases \citep{halloran}. One of the main goal of such simulations is to more accurately reflect the dynamics of the transmission process. For sexually transmitted diseases, learning about the mechanisms of network formation is an important step in that direction.

In this paper, we have used basic ABC and basic regression adjustment techniques because our goal here is to see whether the ABC approach is effective in its simplest and most interpretable form. More refined variants of these methods, which can substantially improve computational performance, can be studied later on. Finally, at the time of writing, we came across related work on how design choices for egocentric network studies impact statistical estimation and inference for ERGMs \citep{krivitsky}. This investigation is relevant for ours, although our focus is specifically on the multiple observation of the evolving network. For a suitably chosen ERGM, i.e., an ERGM with reasonably simple dependence assumptions, it is possible to attain sufficient summary statistics from egocentric network samples. This allows for exact statistical inference, but at the cost of making distributional assumptions that may not hold. For that reason, it is valuable for investigators to have various methods at their disposal so that they may choose the tool that best fits the scientific problem at hand.

\section{Declarations}

\subsection{Availability of data and materials}
Data was simulated using a mechanistic model introduced to study MSM contact networks in Stockholm, Sweden \citep{Hansson}. Our code is accessible at: \href{https://github.com/onnela-lab/longitudinal-inference}{https://github.com/onnela-lab/longitudinal-inference}

\subsection{Competing Interests}
The authors have no competing interest to report. 

\subsection{Funding}
NIH Award \#R01AI138901.

\subsection{Authors' contributions}
All authors conceived the study as well as drafted and revised the manuscript; OS implemented the method in code and carried out data analyses. TH and JP supervised.

\subsection{Acknowledgments}
We would like to acknowledge John Quackenbush and Marcello Pagano for their thoughtful insights on summary statistic exploration.



\bibliography{sn-bibliography}


\begin{thebibliography}{32}
\ifx \bisbn   \undefined \def \bisbn  #1{ISBN #1}\fi
\ifx \binits  \undefined \def \binits#1{#1}\fi
\ifx \bauthor  \undefined \def \bauthor#1{#1}\fi
\ifx \batitle  \undefined \def \batitle#1{#1}\fi
\ifx \bjtitle  \undefined \def \bjtitle#1{#1}\fi
\ifx \bvolume  \undefined \def \bvolume#1{\textbf{#1}}\fi
\ifx \byear  \undefined \def \byear#1{#1}\fi
\ifx \bissue  \undefined \def \bissue#1{#1}\fi
\ifx \bfpage  \undefined \def \bfpage#1{#1}\fi
\ifx \blpage  \undefined \def \blpage #1{#1}\fi
\ifx \burl  \undefined \def \burl#1{\textsf{#1}}\fi
\ifx \doiurl  \undefined \def \doiurl#1{\url{https://doi.org/#1}}\fi
\ifx \betal  \undefined \def \betal{\textit{et al.}}\fi
\ifx \binstitute  \undefined \def \binstitute#1{#1}\fi
\ifx \binstitutionaled  \undefined \def \binstitutionaled#1{#1}\fi
\ifx \bctitle  \undefined \def \bctitle#1{#1}\fi
\ifx \beditor  \undefined \def \beditor#1{#1}\fi
\ifx \bpublisher  \undefined \def \bpublisher#1{#1}\fi
\ifx \bbtitle  \undefined \def \bbtitle#1{#1}\fi
\ifx \bedition  \undefined \def \bedition#1{#1}\fi
\ifx \bseriesno  \undefined \def \bseriesno#1{#1}\fi
\ifx \blocation  \undefined \def \blocation#1{#1}\fi
\ifx \bsertitle  \undefined \def \bsertitle#1{#1}\fi
\ifx \bsnm \undefined \def \bsnm#1{#1}\fi
\ifx \bsuffix \undefined \def \bsuffix#1{#1}\fi
\ifx \bparticle \undefined \def \bparticle#1{#1}\fi
\ifx \barticle \undefined \def \barticle#1{#1}\fi
\bibcommenthead
\ifx \bconfdate \undefined \def \bconfdate #1{#1}\fi
\ifx \botherref \undefined \def \botherref #1{#1}\fi
\ifx \url \undefined \def \url#1{\textsf{#1}}\fi
\ifx \bchapter \undefined \def \bchapter#1{#1}\fi
\ifx \bbook \undefined \def \bbook#1{#1}\fi
\ifx \bcomment \undefined \def \bcomment#1{#1}\fi
\ifx \oauthor \undefined \def \oauthor#1{#1}\fi
\ifx \citeauthoryear \undefined \def \citeauthoryear#1{#1}\fi
\ifx \endbibitem  \undefined \def \endbibitem {}\fi
\ifx \bconflocation  \undefined \def \bconflocation#1{#1}\fi
\ifx \arxivurl  \undefined \def \arxivurl#1{\textsf{#1}}\fi
\csname PreBibitemsHook\endcsname

\bibitem[\protect\citeauthoryear{Macal et~al.}{2004}]{Macal}
\begin{bchapter}
\bauthor{\bsnm{Macal}, \binits{C.}},
\bauthor{\bsnm{Sallach}, \binits{D.}},
\bauthor{\bsnm{North}, \binits{M.}}:
\bctitle{Emergent structures from trust relationships in supply chains}.
In: \bbtitle{Proc. Agent 2004: Conf. on Social Dynamics},
pp. \bfpage{7}--\blpage{9}
(\byear{2004})
\end{bchapter}
\endbibitem

\bibitem[\protect\citeauthoryear{Scholtens and Gentleman}{2005}]{Scholtens}
\begin{botherref}
\oauthor{\bsnm{Scholtens}, \binits{D.}},
\oauthor{\bsnm{Gentleman}, \binits{R.}}:
Making sense of high-throughput protein-protein interaction data.
Statistical Applications in Genetics and Molecular Biology
\textbf{3}(1)
(2005)
\end{botherref}
\endbibitem

\bibitem[\protect\citeauthoryear{Le et~al.}{2022}]{Le}
\begin{barticle}
\bauthor{\bsnm{Le}, \binits{T.-M.}},
\bauthor{\bsnm{Raynal}, \binits{L.}},
\bauthor{\bsnm{Talbot}, \binits{O.}},
\bauthor{\bsnm{Hambridge}, \binits{H.}},
\bauthor{\bsnm{Drovandi}, \binits{C.}},
\bauthor{\bsnm{Mira}, \binits{A.}},
\bauthor{\bsnm{Mengersen}, \binits{K.}},
\bauthor{\bsnm{Onnela}, \binits{J.-P.}}:
\batitle{Framework for assessing and easing global {COVID-19} travel restrictions}.
\bjtitle{Scientific Reports}
\bvolume{12}(\bissue{1}),
\bfpage{1}--\blpage{13}
(\byear{2022})
\end{barticle}
\endbibitem

\bibitem[\protect\citeauthoryear{Adamic and Huberman}{2000}]{WWW}
\begin{barticle}
\bauthor{\bsnm{Adamic}, \binits{L.A.}},
\bauthor{\bsnm{Huberman}, \binits{B.A.}}:
\batitle{Power-law distribution of the world wide web}.
\bjtitle{Science}
\bvolume{287}(\bissue{5461}),
\bfpage{2115}--\blpage{2115}
(\byear{2000})
\end{barticle}
\endbibitem

\bibitem[\protect\citeauthoryear{Robins et~al.}{2007}]{ERGMs}
\begin{barticle}
\bauthor{\bsnm{Robins}, \binits{G.}},
\bauthor{\bsnm{Pattison}, \binits{P.}},
\bauthor{\bsnm{Kalish}, \binits{Y.}},
\bauthor{\bsnm{Lusher}, \binits{D.}}:
\batitle{An introduction to exponential random graph (p*) models for social networks}.
\bjtitle{Social Networks}
\bvolume{29}(\bissue{2}),
\bfpage{173}--\blpage{191}
(\byear{2007})
\end{barticle}
\endbibitem

\bibitem[\protect\citeauthoryear{Goyal and Onnela}{2020}]{ERGMs_Onnela}
\begin{botherref}
\oauthor{\bsnm{Goyal}, \binits{R.}},
\oauthor{\bsnm{Onnela}, \binits{J.}}:
Framework for converting mechanistic network models to probabilistic models.
arXiv 2001.08521
(2020)
\end{botherref}
\endbibitem

\bibitem[\protect\citeauthoryear{Albert and Barab{\'a}si}{2002}]{BA}
\begin{barticle}
\bauthor{\bsnm{Albert}, \binits{R.}},
\bauthor{\bsnm{Barab{\'a}si}, \binits{A.-L.}}:
\batitle{Statistical mechanics of complex networks}.
\bjtitle{Reviews of Modern Physics}
\bvolume{74}(\bissue{1}),
\bfpage{47}
(\byear{2002})
\end{barticle}
\endbibitem

\bibitem[\protect\citeauthoryear{Wertheim et~al.}{2011}]{victor}
\begin{barticle}
\bauthor{\bsnm{Wertheim}, \binits{J.O.}},
\bauthor{\bsnm{Kosakovsky~Pond}, \binits{S.L.}},
\bauthor{\bsnm{Little}, \binits{S.J.}},
\bauthor{\bsnm{De~Gruttola}, \binits{V.}}:
\batitle{Using {HIV} transmission networks to investigate community effects in {HIV} prevention trials}.
\bjtitle{PloS One}
\bvolume{6}(\bissue{11}),
\bfpage{27775}
(\byear{2011})
\end{barticle}
\endbibitem

\bibitem[\protect\citeauthoryear{Aroke et~al.}{2022}]{Opiod_ERGMS}
\begin{bchapter}
\bauthor{\bsnm{Aroke}, \binits{H.}},
\bauthor{\bsnm{Katenka}, \binits{N.}},
\bauthor{\bsnm{Kogut}, \binits{S.}},
\bauthor{\bsnm{Buchanan}, \binits{A.}}:
\bctitle{Network-based analysis of prescription opioids dispensing using exponential random graph models {(ERGMs)}}.
In: \bbtitle{Complex Networks \& Their Applications X: Volume 2, Proceedings of the Tenth International Conference on Complex Networks and Their Applications COMPLEX NETWORKS 2021 10},
pp. \bfpage{716}--\blpage{730}
(\byear{2022}).
\bcomment{Springer}
\end{bchapter}
\endbibitem

\bibitem[\protect\citeauthoryear{Rolls et~al.}{2013}]{PWID}
\begin{barticle}
\bauthor{\bsnm{Rolls}, \binits{D.A.}},
\bauthor{\bsnm{Wang}, \binits{P.}},
\bauthor{\bsnm{Jenkinson}, \binits{R.}},
\bauthor{\bsnm{Pattison}, \binits{P.E.}},
\bauthor{\bsnm{Robins}, \binits{G.L.}},
\bauthor{\bsnm{Sacks-Davis}, \binits{R.}},
\bauthor{\bsnm{Daraganova}, \binits{G.}},
\bauthor{\bsnm{Hellard}, \binits{M.}},
\bauthor{\bsnm{McBryde}, \binits{E.}}:
\batitle{Modelling a disease-relevant contact network of people who inject drugs}.
\bjtitle{Social Networks}
\bvolume{35}(\bissue{4}),
\bfpage{699}--\blpage{710}
(\byear{2013})
\end{barticle}
\endbibitem

\bibitem[\protect\citeauthoryear{Birkett et~al.}{2015}]{chicago}
\begin{barticle}
\bauthor{\bsnm{Birkett}, \binits{M.}},
\bauthor{\bsnm{Armbruster}, \binits{B.}},
\bauthor{\bsnm{Mustanski}, \binits{B.}}, \betal:
\batitle{A data-driven simulation of {HIV} spread among young men who have sex with men: the role of age and race mixing, and {STIs}}.
\bjtitle{Journal of Acquired Immune Deficiency Syndromes}
\bvolume{70}(\bissue{2}),
\bfpage{186}
(\byear{2015})
\end{barticle}
\endbibitem

\bibitem[\protect\citeauthoryear{Mei et~al.}{2010}]{Mei}
\begin{barticle}
\bauthor{\bsnm{Mei}, \binits{S.}},
\bauthor{\bsnm{Sloot}, \binits{P.M.}},
\bauthor{\bsnm{Quax}, \binits{R.}},
\bauthor{\bsnm{Zhu}, \binits{Y.}},
\bauthor{\bsnm{Wang}, \binits{W.}}:
\batitle{Complex agent networks explaining the {HIV} epidemic among homosexual men in {Amsterdam}}.
\bjtitle{Mathematics and Computers in Simulation}
\bvolume{80}(\bissue{5}),
\bfpage{1018}--\blpage{1030}
(\byear{2010})
\end{barticle}
\endbibitem

\bibitem[\protect\citeauthoryear{Hansson et~al.}{2019}]{Hansson}
\begin{barticle}
\bauthor{\bsnm{Hansson}, \binits{D.}},
\bauthor{\bsnm{Leung}, \binits{K.Y.}},
\bauthor{\bsnm{Britton}, \binits{T.}},
\bauthor{\bsnm{Str{\"o}mdahl}, \binits{S.}}:
\batitle{A dynamic network model to disentangle the roles of steady and casual partners for {HIV} transmission among {MSM}}.
\bjtitle{Epidemics}
\bvolume{27},
\bfpage{66}--\blpage{76}
(\byear{2019})
\end{barticle}
\endbibitem

\bibitem[\protect\citeauthoryear{Padeniya}{2021}]{Padeniya}
\begin{botherref}
\oauthor{\bsnm{Padeniya}, \binits{S.M.T.N.}}:
Mathematical modelling to explore the role of the female-sex-worker-client interaction for gonorrhoea transmission and prevention among australian heterosexuals.
PhD thesis,
UNSW Sydney
(2021)
\end{botherref}
\endbibitem

\bibitem[\protect\citeauthoryear{Vajdi et~al.}{2020}]{Vajdi}
\begin{barticle}
\bauthor{\bsnm{Vajdi}, \binits{A.}},
\bauthor{\bsnm{Juher}, \binits{D.}},
\bauthor{\bsnm{Salda{\~n}a}, \binits{J.}},
\bauthor{\bsnm{Scoglio}, \binits{C.}}:
\batitle{A multilayer temporal network model for {STD} spreading accounting for permanent and casual partners}.
\bjtitle{Scientific Reports}
\bvolume{10}(\bissue{1}),
\bfpage{1}--\blpage{12}
(\byear{2020})
\end{barticle}
\endbibitem

\bibitem[\protect\citeauthoryear{Fitzmaurice et~al.}{2012}]{ALA}
\begin{botherref}
\oauthor{\bsnm{Fitzmaurice}, \binits{G.M.}},
\oauthor{\bsnm{Laird}, \binits{N.M.}},
\oauthor{\bsnm{Ware}, \binits{J.H.}}:
Applied longitudinal analysis
(2012)
\end{botherref}
\endbibitem

\bibitem[\protect\citeauthoryear{Csill{\'e}ry et~al.}{2010}]{csillery}
\begin{barticle}
\bauthor{\bsnm{Csill{\'e}ry}, \binits{K.}},
\bauthor{\bsnm{Blum}, \binits{M.G.}},
\bauthor{\bsnm{Gaggiotti}, \binits{O.E.}},
\bauthor{\bsnm{Fran{\c{c}}ois}, \binits{O.}}:
\batitle{Approximate bayesian computation {(ABC)} in practice}.
\bjtitle{Trends in Ecology \& Evolution}
\bvolume{25}(\bissue{7}),
\bfpage{410}--\blpage{418}
(\byear{2010})
\end{barticle}
\endbibitem

\bibitem[\protect\citeauthoryear{Malone et~al.}{2018}]{malone}
\begin{barticle}
\bauthor{\bsnm{Malone}, \binits{J.}},
\bauthor{\bsnm{Syvertsen}, \binits{J.L.}},
\bauthor{\bsnm{Johnson}, \binits{B.E.}},
\bauthor{\bsnm{Mimiaga}, \binits{M.J.}},
\bauthor{\bsnm{Mayer}, \binits{K.H.}},
\bauthor{\bsnm{Bazzi}, \binits{A.R.}}:
\batitle{Negotiating sexual safety in the era of biomedical {HIV} prevention: relationship dynamics among male couples using pre-exposure prophylaxis}.
\bjtitle{Culture, Health \& Sexuality}
\bvolume{20}(\bissue{6}),
\bfpage{658}--\blpage{672}
(\byear{2018})
\end{barticle}
\endbibitem

\bibitem[\protect\citeauthoryear{Down et~al.}{2017}]{down2017}
\begin{barticle}
\bauthor{\bsnm{Down}, \binits{I.}},
\bauthor{\bsnm{Ellard}, \binits{J.}},
\bauthor{\bsnm{Bavinton}, \binits{B.R.}},
\bauthor{\bsnm{Brown}, \binits{G.}},
\bauthor{\bsnm{Prestage}, \binits{G.}}:
\batitle{In {Australia}, most {HIV} infections among gay and bisexual men are attributable to sex with ‘new’partners}.
\bjtitle{AIDS and Behavior}
\bvolume{21}(\bissue{8}),
\bfpage{2543}--\blpage{2550}
(\byear{2017})
\end{barticle}
\endbibitem

\bibitem[\protect\citeauthoryear{de~Vroome et~al.}{2000}]{de2000}
\begin{barticle}
\bauthor{\bsnm{Vroome}, \binits{E.M.}},
\bauthor{\bsnm{Stroebe}, \binits{W.}},
\bauthor{\bsnm{Sandfort}, \binits{T.G.}},
\bauthor{\bsnm{WIT}, \binits{J.B.}},
\bauthor{\bsnm{Griensven}, \binits{G.J.}}:
\batitle{Safer sex in social context: Individualistic and relational determinants of {AIDS}-preventive behavior among gay men 1}.
\bjtitle{Journal of Applied Social Psychology}
\bvolume{30}(\bissue{11}),
\bfpage{2322}--\blpage{2340}
(\byear{2000})
\end{barticle}
\endbibitem

\bibitem[\protect\citeauthoryear{Wall et~al.}{2013}]{wall}
\begin{barticle}
\bauthor{\bsnm{Wall}, \binits{K.M.}},
\bauthor{\bsnm{Stephenson}, \binits{R.}},
\bauthor{\bsnm{Sullivan}, \binits{P.S.}}:
\batitle{Frequency of sexual activity with most recent male partner among young, internet-using men who have sex with men in the {United States}}.
\bjtitle{Journal of Homosexuality}
\bvolume{60}(\bissue{10}),
\bfpage{1520}--\blpage{1538}
(\byear{2013})
\end{barticle}
\endbibitem

\bibitem[\protect\citeauthoryear{Davidovich}{2006}]{davidovich}
\begin{botherref}
\oauthor{\bsnm{Davidovich}, \binits{E.}}:
Liaisons dangereuses: {HIV} risk behavior and prevention in steady gay relationships
(2006)
\end{botherref}
\endbibitem

\bibitem[\protect\citeauthoryear{Weiss et~al.}{2020}]{weiss}
\begin{barticle}
\bauthor{\bsnm{Weiss}, \binits{K.M.}},
\bauthor{\bsnm{Goodreau}, \binits{S.M.}},
\bauthor{\bsnm{Morris}, \binits{M.}},
\bauthor{\bsnm{Prasad}, \binits{P.}},
\bauthor{\bsnm{Ramaraju}, \binits{R.}},
\bauthor{\bsnm{Sanchez}, \binits{T.}},
\bauthor{\bsnm{Jenness}, \binits{S.M.}}:
\batitle{Egocentric sexual networks of men who have sex with men in the {United States}: Results from the {ART}net study}.
\bjtitle{Epidemics}
\bvolume{30},
\bfpage{100386}
(\byear{2020})
\end{barticle}
\endbibitem

\bibitem[\protect\citeauthoryear{Bavinton et~al.}{2016}]{bavinton}
\begin{barticle}
\bauthor{\bsnm{Bavinton}, \binits{B.R.}},
\bauthor{\bsnm{Duncan}, \binits{D.}},
\bauthor{\bsnm{Grierson}, \binits{J.}},
\bauthor{\bsnm{Zablotska}, \binits{I.B.}},
\bauthor{\bsnm{Down}, \binits{I.A.}},
\bauthor{\bsnm{Grulich}, \binits{A.E.}},
\bauthor{\bsnm{Prestage}, \binits{G.P.}}:
\batitle{The meaning of ‘regular partner’in {HIV} research among gay and bisexual men: implications of an australian cross-sectional survey}.
\bjtitle{AIDS and Behavior}
\bvolume{20}(\bissue{8}),
\bfpage{1777}--\blpage{1784}
(\byear{2016})
\end{barticle}
\endbibitem

\bibitem[\protect\citeauthoryear{Myers et~al.}{1999}]{myers}
\begin{botherref}
\oauthor{\bsnm{Myers}, \binits{T.}},
\oauthor{\bsnm{Allman}, \binits{D.}},
\oauthor{\bsnm{Calzavara}, \binits{L.}},
\oauthor{\bsnm{Morrison}, \binits{K.}},
\oauthor{\bsnm{Marchand}, \binits{R.}},
\oauthor{\bsnm{Major}, \binits{C.}}:
Gay and bisexual men's sexual partnerships and variations in risk behaviour
(1999)
\end{botherref}
\endbibitem

\bibitem[\protect\citeauthoryear{Sisson et~al.}{2018}]{Sisson}
\begin{botherref}
\oauthor{\bsnm{Sisson}, \binits{S.A.}},
\oauthor{\bsnm{Fan}, \binits{Y.}},
\oauthor{\bsnm{Beaumont}, \binits{M.}}:
Handbook of approximate bayesian computation
(2018)
\end{botherref}
\endbibitem

\bibitem[\protect\citeauthoryear{Beaumont}{2019}]{Beaumont}
\begin{barticle}
\bauthor{\bsnm{Beaumont}, \binits{M.A.}}:
\batitle{Approximate bayesian computation}.
\bjtitle{Annual Review of Statistics and its Application}
\bvolume{6},
\bfpage{379}--\blpage{403}
(\byear{2019})
\end{barticle}
\endbibitem

\bibitem[\protect\citeauthoryear{Beaumont et~al.}{2002}]{Beaumont2}
\begin{barticle}
\bauthor{\bsnm{Beaumont}, \binits{M.A.}},
\bauthor{\bsnm{Zhang}, \binits{W.}},
\bauthor{\bsnm{Balding}, \binits{D.J.}}:
\batitle{Approximate bayesian computation in population genetics}.
\bjtitle{Genetics}
\bvolume{162}(\bissue{4}),
\bfpage{2025}--\blpage{2035}
(\byear{2002})
\end{barticle}
\endbibitem

\bibitem[\protect\citeauthoryear{Fearnhead and Prangle}{2012}]{Fearnhead}
\begin{barticle}
\bauthor{\bsnm{Fearnhead}, \binits{P.}},
\bauthor{\bsnm{Prangle}, \binits{D.}}:
\batitle{Constructing summary statistics for approximate bayesian computation: semi-automatic approximate bayesian computation}.
\bjtitle{Journal of the Royal Statistical Society: Series B (Statistical Methodology)}
\bvolume{74}(\bissue{3}),
\bfpage{419}--\blpage{474}
(\byear{2012})
\end{barticle}
\endbibitem

\bibitem[\protect\citeauthoryear{Helleringer and Kohler}{2007}]{helleringer}
\begin{barticle}
\bauthor{\bsnm{Helleringer}, \binits{S.}},
\bauthor{\bsnm{Kohler}, \binits{H.-P.}}:
\batitle{Sexual network structure and the spread of {HIV} in {Africa}: evidence from {Likoma Island, Malawi}}.
\bjtitle{Aids}
\bvolume{21}(\bissue{17}),
\bfpage{2323}--\blpage{2332}
(\byear{2007})
\end{barticle}
\endbibitem

\bibitem[\protect\citeauthoryear{Halloran et~al.}{2017}]{halloran}
\begin{barticle}
\bauthor{\bsnm{Halloran}, \binits{M.E.}},
\bauthor{\bsnm{Auranen}, \binits{K.}},
\bauthor{\bsnm{Baird}, \binits{S.}},
\bauthor{\bsnm{Basta}, \binits{N.E.}},
\bauthor{\bsnm{Bellan}, \binits{S.E.}},
\bauthor{\bsnm{Brookmeyer}, \binits{R.}},
\bauthor{\bsnm{Cooper}, \binits{B.S.}},
\bauthor{\bsnm{DeGruttola}, \binits{V.}},
\bauthor{\bsnm{Hughes}, \binits{J.P.}},
\bauthor{\bsnm{Lessler}, \binits{J.}}, \betal:
\batitle{Simulations for designing and interpreting intervention trials in infectious diseases}.
\bjtitle{BMC Medicine}
\bvolume{15}(\bissue{1}),
\bfpage{1}--\blpage{8}
(\byear{2017})
\end{barticle}
\endbibitem

\bibitem[\protect\citeauthoryear{Krivitsky et~al.}{2022}]{krivitsky}
\begin{barticle}
\bauthor{\bsnm{Krivitsky}, \binits{P.N.}},
\bauthor{\bsnm{Morris}, \binits{M.}},
\bauthor{\bsnm{Bojanowski}, \binits{M.}}:
\batitle{Impact of survey design on estimation of exponential-family random graph models from egocentrically-sampled data}.
\bjtitle{Social Networks}
\bvolume{69},
\bfpage{22}--\blpage{34}
(\byear{2022})
\end{barticle}
\endbibitem

\end{thebibliography}

\end{document}